\documentclass[10pt]{article}
\usepackage{epsfig}
\begin{document}

\begin{center}
{\Large \bf ADD extra dimensional gravity and di-muon production at LHC} \\

\vspace{4mm}

I.~Golutvin \footnote{golutvin@sunse.jinr.ru}, \underline{A.~Sapronov} \footnote{sapronov@sunse.jinr.ru},
M.~Savina \footnote{savina@theor.jinr.ru}, S.~Shmatov \footnote{shmatov@cern.ch}\\
\it Laboratory of Particle Physics, \\
\it Joint Institute for Nuclear Research\\

\end{center}


\begin{abstract}
   A possibility to observe TeV-scale gravity signals at the LHC is discussed.
   The ADD scenario with large extra dimensions is considered and its LHC discovery potential
   is derived studying by muon pairs with large invariant masses.

\end{abstract}

\section{Introduction. ADD scenario overview}

Recently several new models of low-scale gravity were proposed based on brane world ideas
\cite{ADD,RS1}. In this work we concentrate on phenomenology of the first of them, the ADD scenario
with flat space-time geometry. We derive the LHC discovery potential to observe one of
several new phenomena appearing in ADD. Namely, we consider modification of the Standard Model
dimuon continuum due to contributions from multiple virtual KK-modes of graviton.

\begin{figure}[ht]
\begin{center}
\resizebox{12cm}{!}{\includegraphics{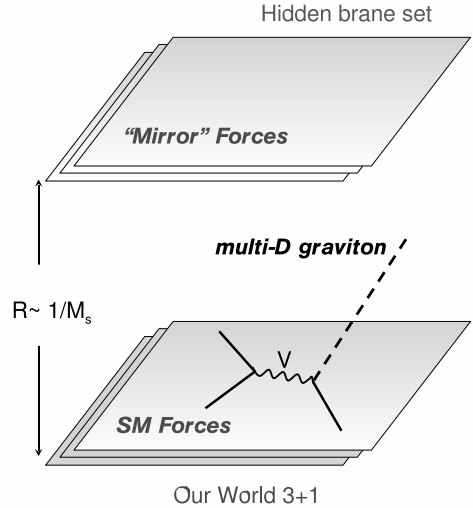}} \vspace{-0.5cm}
\caption{The ADD world with two stacks of branes one of which is
hidden.} \label{Fig1}
\end{center}
\end{figure}

The ADD model \cite{ADD} implies that $n$ ($n$=1..6) extra spatial dimensions can exist in addition to
our three compactified on a $n$-sphere with a radius $R$ (the simlpest case). Then $R$ is called the compactification radius
and it does not have to be the planckian size, but can be really such a large as tenths of millimeter or smaller.
Usual consideration is that all of the Standard Model fields are confined
on a three-brane embedded in a $(3+n)$-dimensional space referred as the bulk (Fig.~\ref{Fig1}). It is the
reason why we don not observe any effects from extra dimensions up to the energy scale $1/R$ when fundamental
multidimensional structure can be distinguished. In this model graviton is only multidimensional field
what can travel through the bulk. In the model the planckian scale $M_{Pl}$ is no longer fundamental but it becomes
the effective scale connected with the true fundamental multidimensional scale $M_S$ in such a way:

\begin{eqnarray}\label{eq:ADD_scale}
    M_{Pl}=M^{1+n/2}_{S}R^{n/2}
\end{eqnarray}
So we can observe possible effects from multidimensional gravity at the energies above $\approx M_{S}$ and
if desired to be probed at the LHC the fundamental mass scale should be adjusted to the order of 1 or
a few TeV.

The characteristic picture of ADD model is the existence of Kaluza-Klein modes of graviton
(these modes will be massive and the mass value is
$m_{KK} = 2\pi k/R$, $k$ is a mode number). These modes must be light: in dependence on a number of extra
dimensions at the fundamental scale $M_{S} \approx 1$ TeV mass values for the first graviton excitation
start from $\approx 10^{-3}$ eV for $n=2$ up to maximal $\approx 10$ MeV for $n=6$.
More details about of ADD phenomenology can be found, for example in \cite{Rubakov01,Kazakov04}.

Setting $M_{S} \approx 1$ TeV one can calculate extra dimensions radius
$R \approx M_{S}^{-1} \times$ $(M_{Pl}/M_{S})^{2/n} \approx$ $10^{(32/n)} \times 10^{-17}$ cm.
The case $n = 1$ gives unacceptably large values of $R$ because the Newton law validity is established down to 0.2 mm \cite{NewtonLaw}.
In the case $n = 2$ the radius value about 1mm, the fundamental scale value $M_{S} \approx 1$ TeV is the most probably excluded
by astrophysics and cosmological arguments (see for example \cite{Rubakov01, ADD2}).
The closest permissible value of $M_{S}$ for $n = 2$ is about 30 TeV that is obviously out of the scope of
observations on modern and future accelerators.
Thus the most favorable set of parameters appears to be $n=3$, $M_{S}\approx1$ TeV and $R \approx 10^{-4}$ mm.

Existing data analysis gave no positive results, but only constraints.
LEP experiments have closely investigated possibilities of presence of large extra dimensions.
One of the evidences would be direct KK-graviton production in $e^+e^- \to \gamma G$ process.
Measurement of final states with photons and missing energy showed no deviation from the Standard
Model predictions and put constraints on $M_{S}$ -- from 1.5 TeV to 0.75 TeV for a number of extra
dimensions from two to five respectively \cite{L3COLL}.

Virtual graviton production at the LEP was searched for in processes with pair-production of fermions
and gauge bosons: $e^+e^- \to e^+e^-$, $\mu^+\mu^-$, $\tau^+\tau^-$, $q\bar{q}$, $\gamma\gamma$,
$ZZ$, $W^+W^-$ \cite{ABBANEO}.
None of the processes exhibited evidence for virtual graviton production and constraints established
for $M_{S}$ are around 1.2 TeV.

TEVATRON experiments CDF and D0 having searched for final states with missing transverse energy in processes
with photons in final state observed no signal.
The CDF collaboration has also searched for events with one or two jets and large missing transverse energy \cite{CDF}.
Virtual graviton production might have been seen as a distortion of mass spectra and angular
distributions of electron and muon pairs as well as diphotons in the final state.
The largest experimental sensitivity was achieved by the combined analysis of electron pairs and
the diphoton channels performed by the D0 collaboration at Run I and Run II.
The data agree with the SM predictions for Drell-Yan electron-pair production, direct diphoton
production and an instrumental background.
The current lower limit on $M_{S}$ derived at TEVATRON are around 1.1 TeV \cite{D0}.

\section{LHC discovery limit}

In the section  above we have pointed out that the characteristic feature of ADD is the existence
of light KK-gravitons which could be directly produced at colliders (real graviton production) or
observed through contact interactions as virtual KK-graviton exchange.
Experimental signals for ADD scenario might be found in dijet, dilepton, diphoton mass spectra
and missing energy distributions.
The missing energy phenomenon corresponds to real graviton production, whereas the first three signals
account for virtual graviton production.
Real gravitons carry away a fraction of the total energy produced in a hard collision, in other words
induce energy leakage from the interaction point.
Virtual gravitons make contribution to the SM diagrams for Drell-Yan processes as well as for gamma
pair production which results in significant modification of these spectra.
An amplitude of each separate graviton contribution is suppressed by $\sim 1/M_{Pl}$, however the production cross
section counts many contributions (large number of gravitons with the same mass value defined by a mode
number $k$, see above, is taken into account with a state density $N(\cal E)$) and this circumstance induces
crucial enhancement of graviton production cross section so effective suppression will be only by $~ (1/M_S)^2$.

The cross section of Drell-Yan process with Kaluza-Klein terms can be factorized as:
\begin{eqnarray}\label{eq:ADD_fact}
    \sigma=\sigma_{SM} + \sigma_4\eta + \sigma_8\eta^2,
\end{eqnarray}
where the first and the third terms correspond to the SM and Kaluza-Klein contributions respectively,
while the second one characterizes an interference between the SM and gravity.
Here $\eta$ is given by
\begin{eqnarray}\label{eq:ADD_eta}
    \eta=\frac{\cal F}{M^4_S}, \qquad
    {\cal F} = \left\{
   \begin{array}{ll}
        \mbox{log}\left( \frac{M^2_S}{\hat s}\right) & for \quad n=2, \\
        \frac{2}{n-2} & for \quad n>2.
    \end{array} \right.
\end{eqnarray}
where $\hat s$ is the center-of-mass energy.
Exact expressions for $\sigma_{SM}$, $\sigma_4\eta$ and $\sigma_8\eta^2$ can be found at
\cite{Cheung03}.

\begin{figure}[ht]
\begin{center}
\hspace{-1cm}
\resizebox{16cm}{!}{\includegraphics{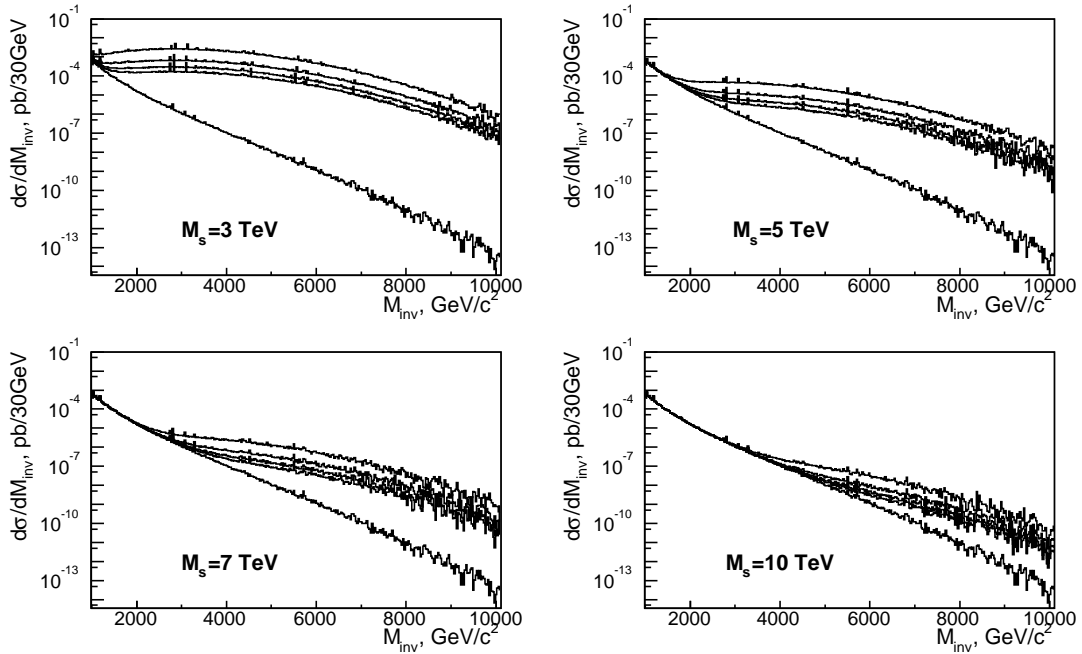}}
\caption{Muon invariant mass for different number
of extra dimensions $n$ (ADD). From bottom to top: SM, $n$ = 6, 5, 4, 3.
Four values of the fundamental gravity scale $M_S$ are considered.}
\label{Fig2}
\end{center}
\vspace{-0.5cm}
\end{figure}
\vspace{0cm}

The characteristic scale value $M_S$ with a number of extra dimensions $n$ pinpoint the region
of dilepton invariant masses $M_{ll}$ sensitive to extra dimensions.
The Fig.~\ref{Fig2} demonstrates the invariant mass distributions of muon pairs for the pure Standard
Model (lower curve) and for scenarios with $n$ = 3, 4, 5, 6 extra dimensions.
Cases with four different values of $M_S$ (3, 5, 7, 10 TeV) were considered.
Those differential cross sections, $d\sigma /dM_{\mu \mu}$, were obtained following
\cite{Cheung03} taking into account the next-leading-order QCD corrections (using the $K$ factor
of 1.38).

In order to estimate the LHC sensitivity to new physics coming from extra dimensions in our analysis we used
typical kinematics and geometrical acceptance of one of the LHC experiments, CMS,
which is, as with ATLAS, expected to be able to trigger and identify
hard muons with a transverse momentum up to several TeV.
The dimuon is accepted when both decay muons are within detector system covering
the pseudorapidity region of $|$${\eta}$$|$ ${\leq}$ 2.4. In
addition, the cut $p_{T}{\geq}20$ GeV/c was applied on each muon.
No cuts were made on isolation of muons in the tracker and the
calorimeter. The total efficiency dimuon selection,
${\varepsilon}$, is about $83{\div}91$ $\%$.
To take into account a detector resolution of a muon momentum the parametrization
$\delta {p_T}\approx 4\%\sqrt{p_T/TeV}$ was used.

The expected significance was computed by method based on counting signal and background events in
a certain  signal region.
The estimator $S_{c12} = 2(\sqrt{N_S + N_B} - \sqrt{N_B})$ \cite{Bityukov00} where $N_S$ and $N_B$
are the number of signal and background events in the invariant mass interval above 1 TeV was
used for this purpose.

\begin{figure}[ht]
\vspace{-1.cm}
\begin{center}
\resizebox{14cm}{!} {\includegraphics{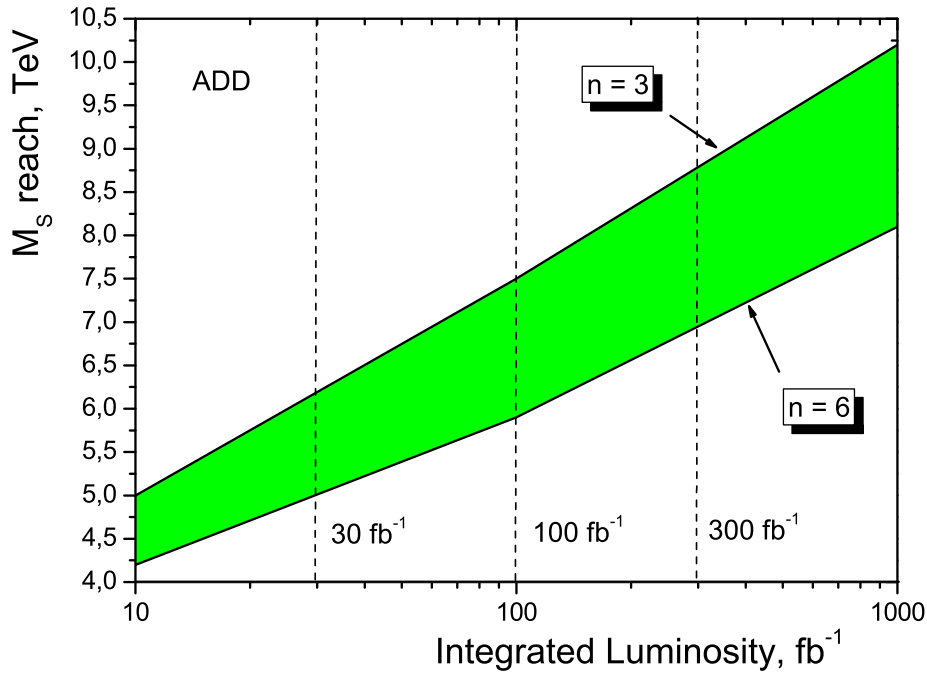}}
\vspace{-0.5cm}
\caption{5$\sigma$ limit on $M_S$ for the number of extra dimensions
$n$ = 3, 4, 5, 6.}
\label{Fig3}
\end{center}
\end{figure}

The ADD model discovery limit as a function of the $M_S$ scale is shown on the Fig.~\ref{Fig3}.
Filled area between two curves ($n$ = 3 and $n$ = 6) shows the upper limit on $M_S$ for different
numbers of extra dimensions with a 5$\sigma$ significance. As it may be seen, the scale $M_S \approx 6 $ TeV can
be reached at integrated luminosity of 100
$\mbox{fb}^{-1}$ even for the most unfavorable case with $n$ = 6.
For more promising scenario where the number of extra dimensions is minimal ($n$ = 3) the
accessible level is extended up to 7.5 TeV.

\vspace{0cm}

Note that these estimates are very close to earlier results \cite{Cheung03} which
were obtained having used the maximum likelyhood method and Bayesian approach.
In this cases the LHC limit extracted from combined dielectron and dimuon analysis is
about 6.9 $\div$ 10.2 TeV for $n$ = 6 $\div$ 2. Generally, the detector performances for
electron and muon measurements are widely different and muon and electron modes
should be analized separetely.

\section{Systematics uncertainties}

As it was discussed above, the physics beyond the Standard Model
can manifest itself as deviations from the standard behavior of
Drell-Yan spectra. At that, these distortions may be both positive
and negative, in other words a dilepton continuum can rise like in
the ADD scenario considered above or equally well fall dawn in
dependence on a theoretical scenario (for instance, in scenario with
non-commutative extra dimensions -- see Ref. \cite{Rubakov01}). And what is more,
a region of dilepton invariant masses where these effects are occurred is
not fixed by a mass window.
It can be more or less narrow resonance state above the dilepton continuum
(like in the RS1 scenario \cite{RS1} or extended gauge models \cite{EGM}), or
just the smooth enough up(down)ward dimuon distribution slope in a
wide invariant mass interval. For correct
extraction of signal events from a background we should understand
reliability of calculations of a dilepton spectrum within the
Standard Model to keep under control all possible sources of
errors and systematic uncertainties. This systematics can be
related to an accuracy of theoretical calculations, an accuracy of
phenomenological determination of PDF's and a roughness of
experimental data -- detector resolution, goodness of fits etc.

A possible theoretical ambiguity in such studies is induced by
incomplete accounting of contributions from QCD and electroweak
higher order quantum corrections to processes considered.
As it was mentioned before current calculations were done
in the leading order for the CTEQ6 with $K$-factor of 1.38 which was used to
take into account next-to-leading-order contributions. It is
expected here that a total value of additional NNLO contributions
does not exceed 5~\%.

A phenomenological origin of PDF gives one another systematic
error. First of all, estimations of a cross section obtained by
using different sets of structure functions are not quite equal.
These results are varying within $\pm 7$~\% for $M_{ll} \ge$ 1
TeV.

In addition, there are uncertainties within the bounds of the same
set (so called {\it internal} uncertainties) coming from an
accuracy of the global analysis of experimental data and from
experimental measurement errors. A recently developed PDF building
technique goes beyond the "standard" paradigm of extracting of
only one "best fit". Last versions of PDF contain a set of various
alternative fits obtained by subjective tuning of specific degrees
of freedom for this PDF \cite{LHAPDF}. Applying standard
statistical methods one can analyze internal
uncertainties comprehensively. These uncertainties are increasing
crucially for very large values of $x$ (or $Q^2$) and in the
small-$x$ region. For instance, uncertainty bands for CTEQ6M set
of PDF's are stayed to be about of 2.6~\%(6~\%) for
$u$($d$)-quarks respectively in the region of $x$ values from
$10^{-3} \div 10^{-4}$ up to 0.3 at $Q^2$ = 10 $\mbox{GeV}^2$, and
these uncertainties grow rapidly for the large $x$ up to of 100
\% at $x$=$0.6 \div 0.7$ \cite{CTEQ}.

\begin{figure}[ht]
\vspace{0cm}
\begin{center}
\resizebox{12cm}{!} {\includegraphics{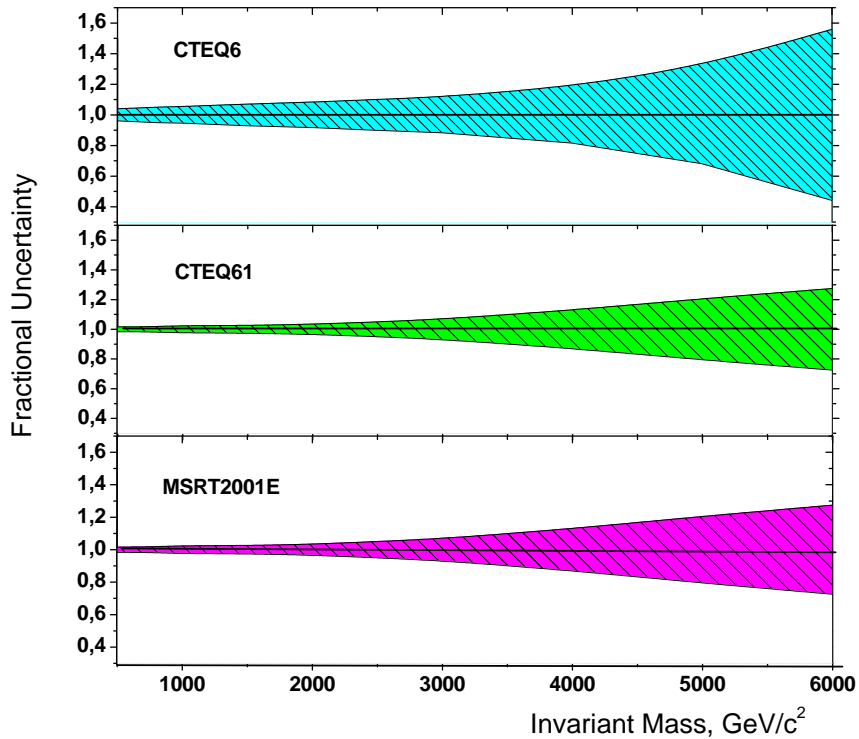}} \vspace{-0.0cm}
\caption{ Uncertainties in calculations of production cross section for
Drell-Yan processes as a function of dimuon invarinat mass.
Cases of different PDF's (CTEQ6, CTEQ61 and MRST2001E) are considered.}
\label{Fig4}
\end{center}
\vspace{0cm}
\end{figure}

Course, an ambiguity in theoretical calculations of the Drell-Yan
cross section due to {\it internal} uncertainties of PDF can also
turn out to be very large. Thus for invariant masses available to
data from the TEVATRON ($\sim 0.6$ TeV) this uncertainty is order
to theoretical one and does not exceed $\pm 6$ \%, but an error
will increase strongly for large values of invariant masses up to
$\sim 10 \div 15$ \% for 3 TeV (Fig.~\ref{Fig4}).

Therefore, PDF related incorrectness of calculations can reduce significantly a
possibility to derive ADD signals. For example, their counting
decreases of the fundamental mass scale reachable at LHC
from 7.5 TeV (for $n$=3) down to 6.5 TeV.
It should be noted that around the region (near 2.5 TeV) the error coming from PDF's is
order to statistical one which is estimated for integrated luminosity of
300 $\mbox{fb}^{-1}$ (about three years of LHC operation in the high-luminosity regime).

Today, the current analysis of PDF sets shows that PDF uncertainties contribute
significantly in Drell-Yan calculation ambiguity.
One expect, that new data from both working
machines, the TEVATRON and HERA, and especially from the LHC in
future will extend an available for analysis range of $x$ and
$Q^2$ and will allow to extract parton distribution functions more
precisely. Moreover, a definite progress is also being expected
from the theoretical side that will be induced by precise counting
both higher twist terms and nonperturbative QCD effects
\cite{CTEQ}.

\section*{Acknowledgements}
Authors would like to thank D.~Bardin, E.~Boos, V.~Rubakov,
and S.~Slabospitsky for enlightening and helpful discussions.
We are very grateful to G.~Landsberg to provide us the ADD generator code.
A.S. would also like to acknowledge specially Organizing Committee for
hospitality and very interesting scientific program of the Conference.

\end{document}